\documentclass[%
 aip,
rsi,%
 amsmath,amssymb,
reprint,%
floatfix,
]{revtex4-1}

\usepackage{graphicx}% Include figure files
\usepackage{dcolumn}% Align table columns on decimal point
\usepackage{braket}
\usepackage{bm}% bold math
\usepackage{multirow}
\usepackage{csquotes}

\usepackage{xcolor}
\usepackage{amsmath}
\usepackage{caption}
\DeclareCaptionType{algorithm}
\usepackage{algpseudocode}

\newcommand{\indnuc}[0]{I}
\newcommand{\indnuctwo}[0]{J}
\newcommand{\then}[1]{N_\mathrm{#1}}

\newcommand{\theone}[0]{\mathbf{1}}

\newcommand{\thet}[0]{t}
\newcommand{\thedt}[0]{\Delta t}

\newcommand{\ther}[1]{\mathbf{R}_{#1}}
\newcommand{\therdot}[1]{\dot{\mathbf{R}}_{#1}}
\newcommand{\therddot}[1]{\ddot{\mathbf{R}}_{#1}}

\newcommand{\thepi}[1]{\boldsymbol{\Pi}_{#1}}
\newcommand{\thepidot}[1]{\boldsymbol{\dot{\Pi}}_{#1}}

\newcommand{\them}[1]{\ifthenelse{\equal{#1}{}}{\mathbf{M}}{M_{#1}}}
\newcommand{\thez}[1]{Z_{#1}}
\newcommand{\theb}[1]{\ifthenelse{\equal{#1}{}}{\mathbf{B}}{B_\mathrm{#1}}}
\newcommand{\thebt}[1]{\ifthenelse{\equal{#1}{}}{\mathbf{\tilde{B}}}{\tilde{B}_{#1}}}
\newcommand{\thef}[2]{\mathbf{F}_{#1}^#2}

\newcommand{\theel}[1]{\mathbf{r}_{#1}}
\newcommand{\theop}[3]{#1_\mathrm{#2}^\mathrm{#3}}

\newcommand{\thewf}[2]{
    \ifthenelse{\equal{#1}{0}}
        {\ifthenelse{\equal{#2}{1}}{\psi_\mathrm{el} (\theel{},\ther{},\thet{})}{\psi_\mathrm{el}}}
        {\ifthenelse{\equal{#2}{1}}{\phi_{#1} (\ther{},\theb{})}{\phi_{#1}}}
}

\newcommand{\thee}[2]{#1_\mathrm{#2}}
\newcommand{\theom}[3]{\boldsymbol{\Omega}_{#1#2} (#3)}
\newcommand{\theq}[3]{Q_{#1#2} (#3)}
\newcommand{\thep}[3]{P_{#1#2} (#3)}

\newcommand{\thew}[3]{\mathbf{W}_{#1#2} (#3)}

\newcommand{\thecoupl}[0]{s}
\newcommand{\thephi}[0]{\chi}

\newcommand{\thecyclofreq}[1]{\omega_{#1}}

\newcommand{\thefaca}[1]{a_{#1}}
\newcommand{\thefacb}[1]{b_{#1}}
\newcommand{\thefacc}[1]{c_{#1}}
\newcommand{\thewt}[1]{\mathbf{w}_{#1}}
\newcommand{\theft}[1]{\mathbf{f}_{#1}}
\newcommand{\theftel}[2]{\mathrm{f}_{#1}^\mathrm{#2}}
\newcommand{\thext}[1]{\mathbf{x}_{#1}}
\newcommand{\thept}[1]{\boldsymbol{\pi}_{#1}}
\newcommand{\theptel}[2]{\mathrm{\pi}_{#1}^\mathrm{#2}}
\newcommand{\thest}[1]{\mathbf{s}_{#1}^{\thecoupl}}
\newcommand{\theut}[2]{\mathbf{u}_{#1}^{#2}}
\newcommand{\theyt}[2]{\mathbf{y}_{#1}^{#2}}
\newcommand{\thevt}[2]{\mathbf{v}_{#1}^{#2}}
\newcommand{\theptdot}[1]{\boldsymbol{\dot{\pi}}_{#1}}
\newcommand{\thextdot}[1]{\mathbf{\dot{x}}_{#1}}

\newcommand{\theconstal}[0]{\alpha}
\newcommand{\theconstbe}[0]{\beta}
\newcommand{\theconstga}[0]{\gamma}

\begin{document}

\title[]{Propagators for molecular dynamics in a magnetic field}% Force line breaks with \\

\author{Laurens D. M. Peters}
\email{laurens.peters@kjemi.uio.no}
\affiliation
{Hylleraas Centre for Quantum Molecular Sciences,  Department of Chemistry, 
University of Oslo, P.O. Box 1033 Blindern, N-0315 Oslo, Norway}
\author{Erik I. Tellgren}
\affiliation
{Hylleraas Centre for Quantum Molecular Sciences,  Department of Chemistry, 
University of Oslo, P.O. Box 1033 Blindern, N-0315 Oslo, Norway}
\author{Trygve Helgaker}
\affiliation
{Hylleraas Centre for Quantum Molecular Sciences,  Department of Chemistry, 
University of Oslo, P.O. Box 1033 Blindern, N-0315 Oslo, Norway}

\date{\today}% It is always \today, today,

\begin{abstract}
\textit{Ab initio} molecular dynamics in a magnetic field requires solving equations of motion with velocity-dependent forces---namely, the Lorentz force arising from the nuclear charges moving in a magnetic field and the Berry force arising from the shielding of these charges from the magnetic field by the surrounding electrons. In this work, we revisit two existing propagators for these equations of motion, the auxiliary-coordinates-and-momenta (ACM) propagator and the Tajima propagator (TAJ), and compare them with a new exponential (EXP) propagator based on the Magnus expansion. Additionally, we explore limits (for example, the zero-shielding limit), the implementation of higher-order integration schemes, and series truncation to reduce  computational cost by carrying out simulations of a HeH$^+$ model system for a wide range of field strengths. While being as efficient as the TAJ propagator, the EXP propagator is the only propagator that converges to both the schemes of Spreiter and Walter (derived for systems without shielding of the Lorentz force) and to the exact cyclotronic motion of a charged particle. Since it also performs best in our model simulations, we conclude that the EXP propagator is the recommended propagator for molecules in magnetic fields. 
\newline
\begin{center}
\includegraphics[width=0.6\textwidth]{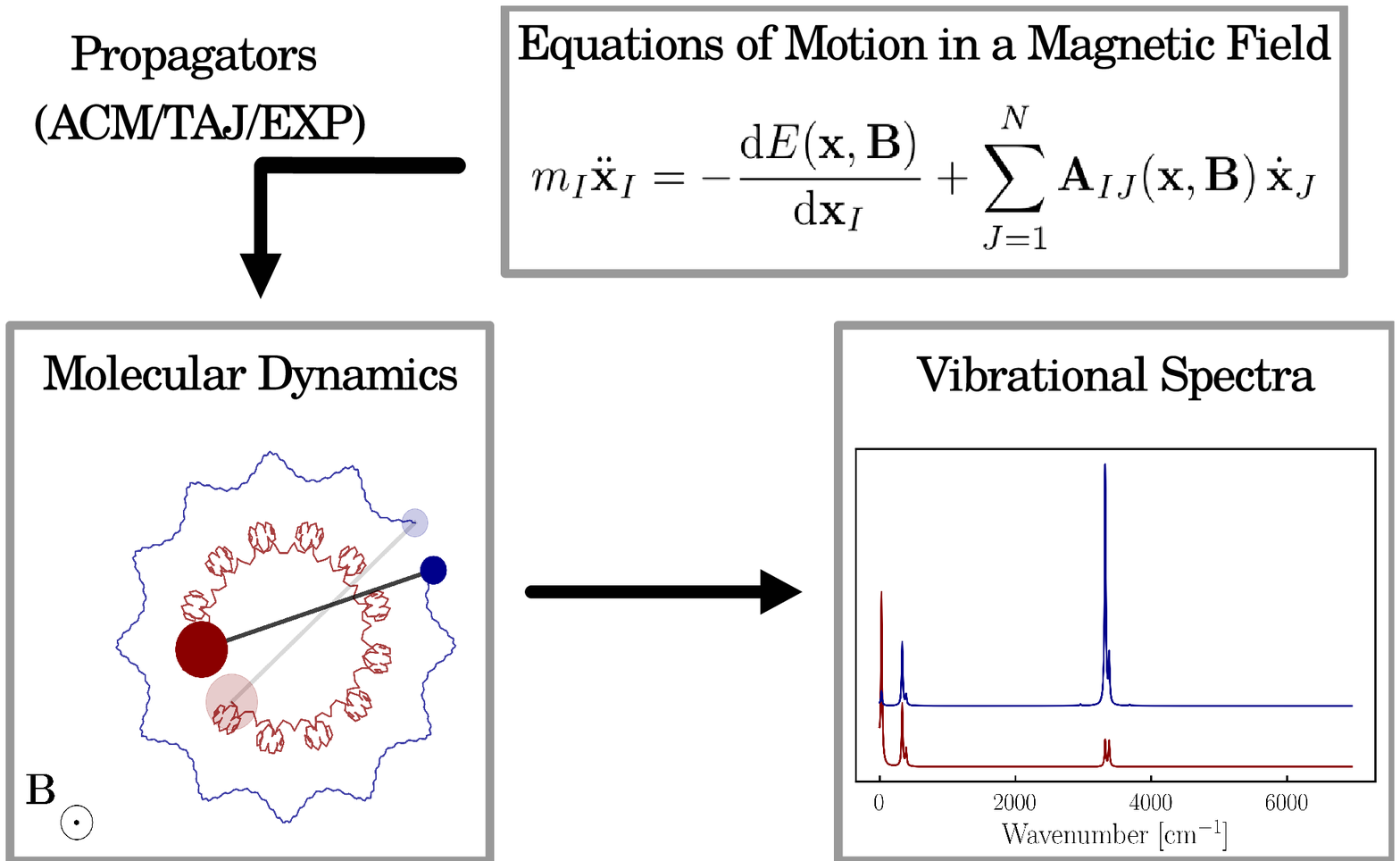}
\end{center}
\par\noindent\rule{0.93\textwidth}{0.1pt}
\end{abstract}

\maketitle

\section{Introduction}

Since the first simulations more than 50 years ago,\cite{Gibson1960,Rahman1964} molecular dynamics has become a ubiquitous tool in computational chemistry, allowing for the calculation of reaction rates and energies,\cite{Chipot} the exploration of reaction networks,\cite{Wang2014,Grimme2019} and the prediction of vibrational spectra,\cite{Futrelle1971,Gaigeot2007,Thomas2013,Thomas2016} including infrared, Raman, and circular dichroism spectroscopies. The principal idea is to solve the equations of motion
\begin{align}
m_I \ddot{\mathbf{x}}_I = - \dfrac{\mathrm{d} E(\mathbf{x})}{\mathrm{d} \mathbf{x}_I} ,\qquad \mathbf{x} = 
\begin{pmatrix}
\mathbf{x}_1 \\
\vdots \\
\mathbf{x}_N 
\end{pmatrix},
\label{int_000}
\end{align}
to determine the motion of a set of $N$ nuclei with masses $m_1, \ldots m_N$ and positions $\mathbf{x}_1(t),\ldots \mathbf{x}_N(t)$. The potential $E(\mathbf{x})$ and the corresponding gradient are usually determined using force-field or (for smaller systems) \textit{ab initio} methods, while the equations of motion are solved using propagators such as the velocity Verlet algorithm\cite{Verlet1967a,Swope1982a} or higher-order schemes.\cite{Omelyan2003a,Blanes2002a} 

In a magnetic field $\mathbf{B}$, a classical particle of charge $q_I$ experiences an additional Lorentz force:
\begin{align}
m_I \ddot{\mathbf{x}}_I = - \dfrac{\mathrm{d} E(\mathbf{x})}{\mathrm{d} \mathbf{x}_I} - q_I \theb{} \times \dot{\mathbf{x}}_I.
\label{int_001}
\end{align}
More than 20 years ago, Spreiter and Walter recognized that the appearance of this velocity-dependent force requires a modification of the velocity Verlet scheme.\cite{Spreiter1999a} Their algorithm, which relies on a Taylor expansion, has been incorporated into well-known software packages\cite{DellaValle2017a,Khajeh2020a} and has been employed in diverse applications.\cite{Al-Haik2006,Chang2006,Daneshvar2020}

Taking into account the electrons as quantum particles within the Born--Oppenheimer approximation, the energy becomes dependent on the magnetic field,\cite{Detmer1997b,Detmer1998b,Tellgren2012a,Lange2012b,Tellgren2014a,Stopkowicz2015b,Austad2020b} while the velocity-dependent force assumes a more general form:\cite{Schmelcher1988d,Ceresoli2007b,Peters2021b}
\begin{align}
m_I \ddot{\mathbf{x}}_I = - \dfrac{\mathrm{d} E(\mathbf{x},\mathbf{B})}{\mathrm{d} \mathbf{x}_I} + 
\sum \limits_{J=1}^{N}
\mathbf{A}_{IJ} (\mathbf{x},\mathbf{B})\, \dot{\mathbf{x}}_J .
\label{int_002}
\end{align}
Here, the last term contains not only  the bare Lorentz force acting on the nuclei [see eq.\,\eqref{int_001}], but also a contribution from the Berry curvature,\cite{Culpitt2021b,Culpitt2022b} reflecting the shielding of the nuclei by the electrons\cite{Peters2022b,Peters2023a} and introducing a coupling between the motion of different nuclei.\cite{Culpitt2023,Tellgren2023} 

While the form and the implications of eq.\,\eqref{int_002} were discussed more than three decades ago,\cite{Schmelcher1988d,Schmelcher1989,Schmelcher1997b}  there are only a few examples where it has been actually solved for a molecular system in a magnetic field. Ceresoli and coworkers simulated an H$_2$ model system in 2007, integrating the equations of motion with a Runge--Kutta scheme.\cite{Ceresoli2007b} In 2021, Peters and coworkers were the first to conduct accurate dynamics of H$_2$ using the auxiliary-coordinates-and-momenta (ACM) propagator,\cite{Peters2021b} which is derived from a general scheme for propagating nonseparable Hamiltonians by Tao.\cite{Tao2016} A more extensive study was conducted by Monzel and coworkers\cite{Monzel2022b} in 2022, studying H$_2$ and LiH with the Tajima (TAJ) propagator, originating from particle physics.\cite{Tajima2018}

In this work, we introduce a new propagator that we refer to as the exponential (EXP) propagator. It is inspired by the fact that equations of type
\begin{align}
m_I \ddot{\mathbf{x}}_I = \mathbf{A}_{II} (\mathbf{x})\, \dot{\mathbf{x}}_I ,
\label{int_003}
\end{align}
appearing, for example, in time-dependent Kohn--Sham theory\cite{GomezPueyo2018} and in surface-hopping algorithms\cite{Barbatti2007}, can be solved exactly using the Magnus expansion\cite{Magnus1954} and approximately using matrix exponential(s) of $\mathbf{A}_{II}$.\cite{Blanes2009} We compare the EXP propagator to the previously published ACM and TAJ propagators, regarding their theoretical foundation, their time-step requirements, their performance during actual simulations, and properties, such as their behaviour in the zero-field case [see eq.~\eqref{int_000}], in the zero-shielding case [see eq.~\eqref{int_001}], and in the cyclotron limit [only the Lorentz force in eq.~\eqref{int_001}]. We use simulations of a HeH$^+$ model system to corroborate our theoretical findings and test schemes to further reduce the computational cost.

Having established the equations of motion, their propagation, and the weak-field limit for the step size in Sections~II.A-C, respectively, we derive the working equations of propagators in the absence of a field as well as the ACM, TAJ, and EXP propagators in Sections~II.D--G and compare them in Section~II.H. Computational details for the HeH$^+$ simulations are given in Section~III, while the results of these simulations are presented and discussed in Section~IV. Conclusions and an outlook are given in Section~V.

\section{Theory}

\subsection{Equations of motion}

Within the Born--Oppenheimer approximation, the equations of motion of a molecule in a magnetic field are given by\cite{Schmelcher1988d,Ceresoli2007b,Peters2021b}
\begin{align}
\them{\indnuc} \therddot{\indnuc} &=  - \dfrac{\partial \thee{E}{}{(\ther{},\theb{})}}{\partial \ther{\indnuc}} \nonumber \\ & \quad +
\sum \limits_{\indnuctwo=1}^{\then{nuc}} 
\left[
\theom{\indnuc}{\indnuctwo}{\ther{},\theb{}} - \delta_{\indnuc\indnuctwo} \thez{\indnuc} \thebt{} 
\right] \therdot{\indnuctwo} .
\label{eom_000}
\end{align}
Here, we use indices $\indnuc{},\indnuctwo{}, ...$ for the $\then{nuc}$ nuclei and $\ther{\indnuc}$, $\thez{\indnuc}$, $\them{\indnuc}$, and $\therdot{\indnuc}$ for the position, charge, mass, and velocity of nucleus $I$. The collective nuclear coordinates are denoted by the $3\then{nuc}$-dimensional column vector $\ther{}$,
\begin{align}
\ther{} = 
\begin{pmatrix}
\ther{1} \\
\vdots \\
\ther{\then{nuc}}
\end{pmatrix} ,
\label{not_003}
\end{align}
while we represent the uniform magnetic field $\theb{}$ of strength $B$ by the $3\times3$ antisymmetric matrix $\thebt{}$ in the manner
\begin{align}
\thebt{} =
\begin{pmatrix} 
0 & - \theb{z} & \theb{y} \\
\theb{z} & 0 & -\theb{x} \\
-\theb{y} & \theb{x} & 0
\end{pmatrix}, \quad \theb{} =
\begin{pmatrix} 
\theb{x} \\
\theb{y} \\
\theb{z} 
\end{pmatrix}, \quad
B = |\theb{}|
\label{not_000}
\end{align}
so that
\begin{align}
\thebt{}\therdot{\indnuc} = 
\theb{} \times \therdot{\indnuc} .
\label{not_001}
\end{align}

The first term in eq.\,\eqref{eom_000} is the gradient of the Born--Oppenheimer electronic energy, obtained at some \textit{ab initio} level of theory,
\begin{align}
\thee{E}{}(\ther{},\theb{}) &= \Braket{\thewf{}{1}|\theop{H}{el}{} (\ther{},\theb{})|\thewf{}{1}} ,
\label{eom_001}
\end{align}
where $\theop{H}{el}{} (\theel{},\ther{},\theb{})$ and $\thewf{}{0} (\theel{};\ther{},\theb{})$ are the electronic Hamiltonian and wave function, respectively, both of which depend on the electronic coordinates $\mathbf r$ [over which the integration is performed in eq.\,\eqref{eom_001}]. The second term in eq.~\eqref{eom_000} is the Lorentz force arising from the nuclear charge screened by the surrounding electrons. This shielding arises from the Berry curvature
\begin{align}
\theom{\indnuc}{\indnuctwo}{\ther{},\theb{}} &= -2 \hbar \Im
\Braket{\dfrac{\partial \thewf{}{1}}{\partial \ther{\indnuc}}|\left[\dfrac{\partial \thewf{}{1}}{\partial \ther{\indnuctwo}}\right]^\mathrm{T}} ,
\label{eom_002}
\end{align}
calculated from derivatives of the electronic wave function with respect to the nuclear coordinates. For more details on the calculation (at the Hartree--Fock level of theory) and interpretation of the Berry curvature, see refs.\,\onlinecite{Culpitt2021b,Culpitt2022b,Peters2022b,Peters2023a}.

With $\them{}$ being the $3\then{nuc}\times3\then{nuc}$-dimensional matrix with the nuclear masses $\them{\indnuc}$ on the diagonal, we can define the kinetic momenta of the nuclei as 
\begin{align}
\thepi{} &= \them{} \therdot{} .
\label{eom_004}
\end{align}
We may now rewrite the nuclear equations of motion in eq.\,\eqref{eom_000} in the more convenient form
\begin{align}
\thepidot{} &= \thef{}{}{} (\ther{},\theb{}) + \thew{}{}{\ther{},\theb{}} \thepi{},
\label{eom_003}
\end{align}
where $\thef{}{}{} (\ther{},\theb{})$ is the $3\then{nuc}$-dimensional vector of the Born--Oppenheimer gradient forces and $\thew{}{}{\ther{},\theb{}}$ is a $3\then{nuc}\times3\then{nuc}$ matrix, whose $\indnuc\indnuctwo$ blocks of dimension $3\times3$ contain the Berry curvature as well as the contribution from the bare (unscreened) Lorentz force:
\begin{align}
\thew{\indnuc}{\indnuctwo}{\ther{},\theb{}} &= 
\them{\indnuctwo}^{-1} \left[\theom{\indnuc}{\indnuctwo}{\ther{},\theb{}} 
- \delta_{\indnuc\indnuctwo}
\thez{\indnuc} \thebt{} \right].
\label{eom_005b}
\end{align}
The form of the equations of motion given in eq.\,\eqref{eom_003} highlights that the effect of the magnetic field is twofold: It changes the potential energy surface leading to different Born--Oppenheimer gradient forces and it introduces an additional, velocity-dependent term. 

\subsection{Propagators: general considerations}

The main objective of this theory section is to discuss how these modified equations of motion can be integrated efficiently using different propagators. We denote by
\begin{align}
\thext{\thefaca{}} &= \ther{}(\thet{} + \thefaca{}\thedt{}) ,\label{eom_006a}\\
\thept{\thefaca{}} &= \them{}\therdot{}(\thet{} + \thefaca{}\thedt{}), \label{eom_006b}
\end{align}
the position and momentum, respectively, of the system at time $ \thefaca{}$ relative to time $\thet{}$ in units of the time step $\thedt{}$. Introducing the notation
\begin{align}
\theft{\thefaca{}} &= \thef{}{}{} (\thext{\thefaca{}},\theb{}) ,\label{eom_006c}\\
\thewt{\thefaca{}} &= \thew{}{}{\thext{\thefaca{}},\theb{}},\label{eom_006d}
\end{align}
we may now express the equations of motion in terms of differentials with respect to the factor $\thefaca{}$
\begin{align}
\theptdot{\thefaca{}} &= 
\thedt{}^{-1} \dfrac{\partial \thept{\thefaca{}}}{\partial \thefaca{}}
= \theft{\thefaca{}} + \thewt{\thefaca{}} \thept{\thefaca{}} ,
\label{std_002a}\\
\thextdot{\thefaca{}} &= 
\thedt{}^{-1} \dfrac{\partial \thext{\thefaca{}}}{\partial \thefaca{}}
= \them{}^{-1} \thept{\thefaca{}} .
\label{std_002b}
\end{align}
The choice of the step length $\thedt{}$, here appearing as a prefactor, will be discussed in the next subsection. 

In this notation, a propagator is defined as a scheme that updates the coordinates ($\thext{0}\rightarrow\thext{1}$) and momenta ($\thept{0}\rightarrow\thept{1}$) for a given $\thedt{}$. Since eqs.~\eqref{std_002a} and \eqref{std_002b} depend on each other, they are usually solved alternately for a series of substeps. To ease their reading, we will only derive working equations for a single set of those substeps,
\begin{align}
\thept{0} \rightarrow \thept{\thefaca{}}: \quad\thept{\thefaca{}} &= \Delta t \! \int \limits_{0}^{\thefaca{}}\! \!(\theft{\theconstal{}}+ \thewt{\theconstal{}}\thept{\theconstal{}})\,\mathrm{d}\theconstal{} + \thept{0} ,
\label{std_001a}\\
\thext{0} \rightarrow \thext{\thefacb{}}: \quad
\thext{\thefacb{}} &= \Delta t\,\them{}^{-1} \int \limits_{0}^{\thefacb{}}\!  \!\thept{\theconstbe{}}\,\mathrm{d}\theconstbe{}
+ \thext{0} ,
\label{std_001b}
\end{align}
where $\thext{0}$ and $\thept{0}$ are initial values and $\thefaca{}$ and $\thefacb{}$ are arbitrary step lengths in units of $\thedt{}$. Equations for all other substeps can then be derived by adjusting the initial values and step lengths accordingly.

To evaluate the integrals in eqs.\,\eqref{std_001a} and \eqref{std_001b}, we will draw inspiration from the mean-value theorems. They state that, for real-valued functions $\varphi(\alpha)$ and $\psi(\alpha)$ that are continuous in $[0,a]$, there exists a point $0 \leq b \leq a$ such that 
\begin{equation}
\int \limits_0^a \!\varphi(\alpha) \, \mathrm d\alpha = a \varphi(b),  \label{mvt0a}
\end{equation}
and
\begin{equation}
\int \limits_0^a \! \varphi(\alpha) \psi(\alpha) \, \mathrm d t = \psi(0)\! \! \int \limits_0^b \!\!\varphi(\alpha) \, \mathrm d\alpha + \psi(a)\!\!\int \limits_b^a \!\!\varphi(\alpha)\, \mathrm d\alpha. \label{mvt0b}
\end{equation}
The corresponding statements are not guaranteed to hold for vector-valued functions; however, we will at one point rely on them as approximations. Specifically, we use the forms
\begin{align}
\int \limits_{0}^{\thefaca{}}\! \! \,\boldsymbol{\varphi} (\theconstal{}) \, \mathrm{d}\theconstal{} &\approx
\boldsymbol{\varphi} (\thefacb{}) \!\! \int \limits_{0}^{\thefaca{}}\! \mathrm{d}\theconstal{} 
= \thefaca{}\boldsymbol{\varphi} (\thefacb{}) ,
\label{mvt}
\end{align}
and
\begin{align}
\int \limits_{0}^{\thefaca{}}\! \!\boldsymbol{\psi} (\theconstal{}) \boldsymbol{\varphi}(\theconstal{}) \,\mathrm{d}\theconstal{} &\approx
\int \limits_{0}^{\thefacb{}}\!\! \boldsymbol{\psi} (\theconstal{}) \boldsymbol{\varphi} (0)\, \mathrm{d}\theconstal{} +
\int \limits_{\thefacb{}}^{\thefaca{}}\! \! \boldsymbol{\psi} (\theconstal{}) \boldsymbol{\varphi} ({\thefaca{}})\,\mathrm{d}\theconstal{} ,
\label{mvt2}
\end{align}
to approximate integrals over two vectors of continuous functions $\boldsymbol{\varphi} (\theconstal{})$ and $\boldsymbol{\psi} (\theconstal{})$. Note that both the midpoint rule and the trapezoidal rule can be derived from these mean-value approximations by choosing $\thefacb{} = \thefaca{}/2$ and, in case of latter, $\boldsymbol{\psi} (\theconstal{}) = \theone{}$:
\begin{align}
\int \limits_{0}^{\thefaca{}}\!  \!\boldsymbol{\varphi} (\theconstal{}) \,\mathrm{d}\theconstal{}&\approx
\boldsymbol{\varphi} (\thefaca{}/2) \!\!\int \limits_{0}^{\thefaca{}}\! \mathrm{d}\theconstal{} 
=
\thefaca{}\boldsymbol{\varphi} (\thefaca{}/2)
\label{mpr} \\
\int \limits_{0}^{\thefaca{}}\! \!\boldsymbol{\varphi} (\theconstal{})\,\mathrm{d}\theconstal{}  &\approx
\int \limits_{0}^{\thefaca{}/2}\!  \!\boldsymbol{\varphi} (0)\,\mathrm{d}\theconstal{} + \int \limits_{\thefaca{}/2}^{\thefaca{}}\!  \!\boldsymbol{\varphi} (1) \, \mathrm{d}\theconstal{}=
\dfrac{\thefaca{}}{2} \left[\boldsymbol{\varphi} (0) + \boldsymbol{\varphi} (1) \right]
\label{tpr}
\end{align}

\subsection{Choice of step size}

The error introduced by a given propagator depends on the applied step size $\thedt{}$. Before considering the different propagators, we consider briefly in this subsection the restriction on the time step imposed by an external magnetic field. As in field-free simulations, we demand that the time step $\thedt{}$ is significantly smaller than the fastest molecular vibration. In addition, we demand that the matrix $\thedt{} \,\thewt{a}$ is convergent for all $a$, meaning that
\begin{align}
\lim \limits_{n \rightarrow{} \infty}  \left [\thedt{} \thewt{a}\right]^{n} = \mathbf{0} 
\label{dt_000}
\end{align}
Introducing the spectral radius $\rho$
\begin{align}
\sqrt{\rho\left( \thewt{a}^\mathrm{T} \thewt{a} \right)} = \thecyclofreq{a} ,
\label{dt_001}
\end{align}
which we loosely interpret as a cyclotron frequency, we have
\begin{align}
\lim \limits_{n \rightarrow{} \infty}  \left\| \left[\thedt{} \thewt{a}\right]^{n} \right\|& \propto  \lim \limits_{n \rightarrow{} \infty} \thedt{}^n \, \rho\left( \thewt{a}^\mathrm{T} \thewt{a} \right)^{n/2} \nonumber \\ &= \lim \limits_{n \rightarrow{} \infty} (\thedt{} \thecyclofreq{a})^n \label{dt_001a}
\end{align}
in some matrix norm $\|\cdot\|$. Consequently, eq.~\eqref{dt_000} holds when 
\begin{align}
\thedt{} < \thecyclofreq{a}^{-1} .
\label{dt_002}
\end{align}
This \emph{weak-field limit}, as defined by Spreiter and Walter,\cite{Spreiter1999a} means that the time step is small enough to resolve the cyclotronic motion. For a neutral molecule, the charge--mass ratio is roughly on the order of $10^{-4}\,$a.u., so that
\begin{align}
 \thecyclofreq{a} \sim 10^{-4} B .
\label{dt_003}
\end{align} 
Time steps of up to 100 a.u.\ (2.4\,fs) therefore allow for simulations in field strengths up to about $10\,\mathrm{B}_0$, covering the entire range of field strengths from the (particularly interesting) intermediate regime ($<1\,\mathrm{B}_0$) up to the Landau regime.

\subsection{Propagators in the absence of a field}

For the field-free case, $\thewt{}$ is zero so that we can directly apply the mean-value approximation [eq.~\eqref{mvt}] to eqs.~\eqref{std_001a} and \eqref{std_001b}:
\begin{align}
\thept{\thefaca{}} &= \Delta t \! \int \limits_{0}^{\thefaca{}}\!  \!\theft{\theconstal{}}\,\mathrm{d}\theconstal{} + \thept{0} 
\approx \thefaca{}\thedt{}\,\theft{\thefacb{}} + \thept{0} 
\label{std_001c}\\
\thext{\thefacb{}} &= \Delta t\,\them{}^{-1} \!\int \limits_{0}^{\thefacb{}}\! \!\thept{\theconstbe{}}\,\mathrm{d}\theconstbe{} 
+ \thext{0} 
\approx 
\thefacb{} \thedt{} \,\them{}^{-1} \, \thept{\thefaca{}} + \thext{0} 
\label{std_001d}
\end{align}
As shown in Algorithm~\ref{alg_std}, both equations are solved alternately using the current forces and momenta as mean values to propagate the momenta and positions, respectively. The remaining task is now to come up with a series of $\thefaca{}$'s and $\thefacb{}$'s (denoted by the vectors $\mathbf{a}$ and $\mathbf{b}$) to approximate the mean values. The lengths of $\mathbf{a}$ and $\mathbf{b}$ ($K+1$ and $K$) reflect the order $K$ of the scheme.

In the standard velocity Verlet scheme ($K = 1$),\cite{Verlet1967a,Swope1982a} we set $\mathbf{a} = [0.5,0.5]$ and $\mathbf{b} = [1.0]$, so that we obtain:
\begin{align}
\thept{0.5}^\mathrm{VV} &= 0.5\thedt\,\theft{0} + \thept{0} 
\label{std_005a}\\
\thext{1}^\mathrm{VV} &= \thedt
\left[
\them{}^{-1} \thept{0.5}^\mathrm{VV}
\right]
+ \thext{0} 
\label{std_005b} \\
\thept{1}^\mathrm{VV} &= 0.5\thedt\,\theft{1} + \thept{0.5}^\mathrm{VV}
\label{std_005c}
\end{align}
Note that this quadrature corresponds to solving the full integral ($\thefaca{}=1$) in eq.~\eqref{std_001c} with the trapezoidal [see eq.~\eqref{tpr}] and the full integral in eq.~\eqref{std_001d} ($\thefacb{}=1$) with the midpoint rule [see eq.~\eqref{mpr}]. As a consequence of this the velocity Verlet algorithm is symplectic, time-reversible, and of second-order accuracy. However, it has been shown that higher-order schemes,\cite{Omelyan2003a,Blanes2002a} can significantly increase the stability of the dynamics. 

Unfortunately, there is no straightforward expansion of the upper scheme towards systems with velocity-dependent forces since, taking the velocity Verlet algorithm as an example, the step of eq.~\eqref{std_005c} becomes
\begin{align}
\thept{1} &\approx 0.5\thedt\left(\theft{1} + \thewt{1}\thept{1}\right)+ \thept{0.5} 
\label{std_006}
\end{align}
The mismatch between the required ($\thept{1}$) and the available ($\thept{0.5}$) momenta for the propagation leads to a systematic error in the integration of the equations of motion.\cite{Peters2021b} Clearly, there is a need to develop alternative propagators, which will be done in the following subsections. 

\begin{algorithm}
\caption{General algorithm for a propagator of order $K$, in the absence of a field. See eqs.~\eqref{eom_006a}-\eqref{eom_006d} for definitions of the variables. $\mathbf{a}$ and $\mathbf{b}$ are predefined vectors of length $K+1$ and $K$, respectively. In case of velocity Verlet, $\mathbf{a} = [0.5,0.5]$ and $\mathbf{b} = [1.0]$.}
\begin{algorithmic}
\State $k = 0$; $a' = 0$; $b' = 0$
\While{$k \leq K$}
  \State $\thefaca{} = \mathbf{a}[k]$
  \State $\thept{a' + \thefaca{}} = \thefaca{}\thedt{} \theft{b'} +  \thept{a'}$
  \State $a' = a' + a$
  \If{$k < K$}
    \State $\thefacb{} = \mathbf{b}[k]$
    \State $\thext{b' + \thefacb{}} =  \thefacb{}\thedt{} \left[ \them{}^{-1} \thept{a'} \right] + \thext{b'} $
    \State $b' = b' + b$
    \State \textbf{Calculate} $\theft{b'}$
  \EndIf
  \State $k = k+1$
\EndWhile
\end{algorithmic}
\label{alg_std}
\par\noindent\rule{0.48\textwidth}{0.1pt}
\end{algorithm}
\newpage
\subsection{Auxiliary-coordinates-and-momenta propagator}

The auxiliary-coordinates-and-momenta (ACM) propagator method was proposed by Tao for general nonseparable Hamiltonians\cite{Tao2016} and adapted by Peters and coworkers for molecular simulations in a magnetic field.\cite{Peters2021b} The idea is to circumvent the mismatch in eq.~\eqref{std_006}, by introducing an additional pair of coordinates and momenta ($\thext{}',\thept{}'$) that are kept close to the original pair ($\thext{},\thept{}$) during the dynamics ($\thext{} \approx \thext{}'$ and $\thept{} \approx \thept{}'$). If this coupling is sufficiently strong, we can write the propagation of $\thept{}$ in terms of $\thept{}'$ and $\thext{}$
\begin{align}
\thept{\thefaca{}} &\approx  \thefaca{} \thedt{} \left[ 
\theft{\thefacb{}} + \thewt{\thefacb{}} \thept{\thefacb{}}' \right] + \thept{0}, \label{aux_001a}\\
\thext{\thefacb{}} &\approx  \thefacb{} \thedt{} \left[ \them{}^{-1} \thept{\thefaca{}} \right] + \thext{0} . \label{aux_002b}
\end{align}
and the propagation of $\thept{}'$ in terms of $\thept{}$ and $\thext{}'$
\begin{align}
\thept{\thefacb{}}' &\approx \thefacb{} \thedt{} \left[ 
\theft{\thefaca{}}' + \thewt{\thefaca{}}' \thept{\thefaca{}} \right] + \thept{0}', \label{aux_002a} \\
\thext{\thefaca{}}' &\approx  \thefaca{} \thedt{} \left[ \them{}^{-1} \thept{\thefacb{}}' \right] + \thext{0}', \label{aux_001b} 
\end{align}
The only remaining step is the coupling of coordinates and momenta, which is done when all quantities $\thext{}$, $\thept{}$, $\thext{}'$, and $\thept{}'$ are at the same time step by carrying out, for a given coupling constant $s$, the transformation\cite{Peters2021b}
\begin{align}
&\begin{pmatrix} 
\thext{\thefacc{}} \\
\thext{\thefacc{}}' \\
\them{}^{-1} \thept{\thefacc{}} \\
\them{}^{-1} \thept{\thefacc{}}'
\end{pmatrix} \rightarrow \thest{\thefacb{}} 
\begin{pmatrix} 
\thext{\thefacc{}} \\
\thext{\thefacc{}}' \\
\them{}^{-1} \thept{\thefacc{}} \\
\them{}^{-1} \thept{\thefacc{}}'
\end{pmatrix} 
\label{aux_003}
\end{align}
with the coupling matrix
\begin{align}
\thest{\thefacb{}} &= \dfrac{1}{2}
\begin{pmatrix} 
1 + \cos{\thephi} & 1 - \cos{\thephi} & \thecoupl{}^{-1} \sin{\thephi} & -\thecoupl{}^{-1} \sin{\thephi} \\
1 - \cos{\thephi} & 1 + \cos{\thephi} & -\thecoupl{}^{-1} \sin{\thephi} & \thecoupl{}^{-1} \sin{\thephi} \\
\thecoupl{}\sin{\thephi} & -\thecoupl{}\sin{\thephi} & 1 + \cos{\thephi} & 1 - \cos{\thephi} \\
-\thecoupl{}\sin{\thephi} & \thecoupl{}\sin{\thephi} & 1 - \cos{\thephi} & 1 + \cos{\thephi}
\end{pmatrix}
\label{aux_004}
\end{align}
where
\begin{align}
\thephi &= \thecoupl{} \thefacb{} \thedt{}
\label{aux_005}
\end{align}
Setting $\thecoupl{} = 0$ reduces the coupling matrix to the unity matrix, while setting it to $\thecoupl{} = \pi/(\thefacc{} \thedt{})$ leads to a complete exchange of $\thext{}$ and $\thept{}$ with $\thext{}'$ and $\thept{}'$ and vice versa. The pseudo code for a general ACM propagator of order $K$ is given in Algorithm~\ref{alg_aux}. 

\begin{algorithm}
\caption{General algorithm for the auxiliary coordinates and momenta (ACM) method of order $K$. See eqs.~\eqref{eom_006a}-\eqref{eom_006d} for definitions of the variables. $\mathbf{a}$ and $\mathbf{b}$ are predefined vectors of length $K+1$ and $K$, respectively. In the velocity Verlet variant, $\mathbf{a} = [0.5,0.5]$ and $\mathbf{b} = [1.0]$.}
\begin{algorithmic}
\State $k = 0$; $a' = 0$; $b' = 0$
\While{$k \leq K$}
  \State $\thefaca{} = \mathbf{a}[k]$
  \State $\thept{a' + \thefaca{}} =  \thefaca{} \thedt{} \left[\theft{b'} + \thewt{b'} \thept{b'}' \right] + \thept{a'}$
  \State $\thext{a' + \thefaca{}}' =  \thefaca{} \thedt{} \left[ \them{}^{-1} \thept{b'}' \right] + \thext{a'}'$
  \State $a' = a' + a$
  \If{$k < K$}
    \State $\thefacb{} = \mathbf{b}[k]$
    \State \textbf{Calculate} $\theft{a'}'$ and $\thewt{a'}'$
    \State  $\thept{a'}' =  (a' - b')\thedt{} \left[ \theft{a'}' + \thewt{a'}' \thept{a'} \right] + \thept{b'}'$
    \State $\thext{a'} =  (a' - b')\thedt{} \left[ \them{}^{-1} \thept{a'} \right] + \thext{b'} $
    \State \textbf{Apply} $\thest{\thefacb{}}$
    \State \textbf{Calculate} $\theft{a'}'$ and $\thewt{a'}'$
    \State $b' = b' + b$
    \State  $\thept{b'}' =  (b' - a')\thedt{} \left[ \theft{a'}' + \thewt{a'}' \thept{a'} \right] + \thept{a'}'$
    \State $\thext{b'} =  (b' - a')\thedt{} \left[ \them{}^{-1} \thept{a'} \right] + \thext{a'}  $
    \State \textbf{Calculate} $\theft{b'}$ and $\thewt{b'}$
  \EndIf
  \State $k = k+1$
\EndWhile
\end{algorithmic}
\label{alg_aux}
\par\noindent\rule{0.48\textwidth}{0.1pt}
\end{algorithm}

\subsection{Tajima propagator}

An alternative to the ACM propagator is the Tajima (TAJ) propagator. Initially used in particle physics,\cite{Tajima2018} it was rewritten for molecular applications by Monzel \emph{et al}.\cite{Monzel2022b} Here, we present a slightly different derivation of the working equations that starts with applying the mean-value approximations [eqs.~\eqref{mvt} and \eqref{mvt2}] to eq.~\eqref{std_001a}:
\begin{align}
\thept{\thefaca{}} &\approx \thefaca{} \thedt{}\,\theft{\thefacb{}} + 
\thedt{} \!\! \int \limits_{0}^{\thefacc{}}\! \!\thewt{\theconstal{}}\thept{0}
\,\mathrm{d}\theconstal{}+
\thedt{} \!\! \int \limits_{\thefacc{}}^{\thefaca{}}\! \!\thewt{\theconstal{}}\thept{\thefaca{}}\,\mathrm{d}\theconstal{} 
 + \thept{0} 
\label{taj_000}
\end{align}
Assuming that $\thefacc{} = \thefaca{}/2$ and that both integrals have the \emph{same} mean value $\thewt{\thefacb{}}$, we obtain:
\begin{align}
\left[
\theone - \dfrac{\thefaca{}}{2} \thedt{}\,\thewt{\thefacb{}}
\right]
\thept{\thefaca{}}
&\approx
\thefaca{} \thedt{}\,\theft{b} + 
\left[
\theone + \dfrac{\thefaca{}}{2} \thedt{}\,\thewt{\thefacb{}}
\right]
\thept{0}
\label{taj_002}
\end{align}
Introducing the inverted matrix
\begin{align}
\thevt{b}{\thefacc{}} &= 
\left [\theone{} - \thefacc{} \thedt{} \thewt{b}\right]^{-1} ,
 \label{taj_003}
\end{align}
we arrive at the working equation for the TAJ propagator, 
\begin{align}
\thept{\thefaca{}} 
&\approx 
\thevt{b}{\thefaca{}/2} \left(\thefaca{} \thedt{} \theft{b} +
 \left[\theone{} + \dfrac{\thefaca{}}{2} \thedt{} \thewt{b}\right] 
\thept{0}
\right) 
 \label{taj_005},
 \end{align}
yielding the algorithm in Algorithm~\ref{alg_taj}. In the \emph{weak-field limit} [see eq.~\eqref{dt_000}], we can expand the matrix inversion in a Neumann series:
\begin{align}
\thevt{b}{\thefacc{}} &= 
\sum \limits_{n=0}^{\infty} \left [\thefacc{} \thedt{} \thewt{b}\right]^{n}  =
\sum \limits_{n=0}^{\then{}-1} \left [\thefacc{} \thedt{} \thewt{b}\right]^{n} + \mathcal{O} \left( 
\left [\thedt{} \thecyclofreq{b} \right]^{\then{}}
\right) 
 \label{taj_004}
\end{align}
In standard applications, we expect this series to converge fast. Using the estimate in eq.~\eqref{dt_003} with $B= 0.1$ and $\thedt{} = 10$ in atomic units, the error is on the order of $10^{-8}$ when using $\then{} = 2$. Consequently, we can avoid the (comparably) high computational cost of matrix inversions during the simulations.

\begin{algorithm}
\caption{General algorithm for the Tajima (TAJ) method of order $K$. See eqs.~\eqref{eom_006a}-\eqref{eom_006d} for definitions of the variables. $\mathbf{a}$ and $\mathbf{b}$ are predefined vectors of length $K+1$ and $K$, respectively. In the velocity Verlet variant, $\mathbf{a} = [0.5,0.5]$ and $\mathbf{b} = [1.0]$.}
\begin{algorithmic}
\State $k = 0$; $a' = 0$; $b' = 0$
\While{$k \leq K$}
  \State $\thefaca{} = \mathbf{a}[k]$
  \State $\thept{a' + \thefaca{}} = \thevt{b'}{\thefaca{}/2} \left(\thefaca{}\thedt{} \theft{b'} +  \left[\theone{} + \dfrac{\thefaca{}}{2} \thedt{} \thewt{b'}\right] \thept{a'}\right)$
  \State $a' = a' + a$
  \If{$k < K$}
    \State $\thefacb{} = \mathbf{b}[k]$
    \State $\thext{b' + \thefacb{}} =   \thefacb{}\thedt{} \left[ \them{}^{-1} \thept{a'} \right] + \thext{b'} $
    \State $b' = b' + b$
    \State \textbf{Calculate} $\theft{b'}$ and $\thewt{b'}$
  \EndIf
  \State $k = k+1$
\EndWhile
\end{algorithmic}
\label{alg_taj}
\par\noindent\rule{0.48\textwidth}{0.1pt}
\end{algorithm}

\subsection{Exponential propagator}

A general approach to solving first-order linear differential equations is the Magnus integrator method. Specifically, we may write
\begin{align}
\thept{\thefaca{}} &= \thedt{} \! \int  \limits_{\theconstga}^{\thefaca{}} \! \exp\left(\theyt{\theconstal{},\thefaca{}}{}\right) \theft{\theconstal{}} \, \mathrm{d} \theconstal + \exp\left(\theyt{\theconstga,\thefaca{}}{}\right) \thept{\theconstga} 
\label{exp_002}
\end{align}
where $\theyt{\theconstal{},\thefaca{}}{}$ is the Magnus series,\cite{Magnus1954}
\begin{align}
\theyt{\theconstga,\thefaca{}}{} & =
\thedt{} \!\int \limits_{\theconstga}^{\thefaca{}} \! \mathrm{d}\theconstal{} \,\thewt{\theconstal{}} \nonumber \\ & \quad +
\dfrac{1}{2} (\thedt{})^2
\int \limits_{\theconstga}^{\thefaca{}} \!
\mathrm{d}\theconstal{} \!
\int \limits_{\theconstga}^{\theconstal} \!
\mathrm{d}\theconstbe{}
\left[
\thewt{\theconstal{}}, \thewt{\theconstbe{}}
\right]
+ \cdots .
\label{exp_000c}
\end{align}
From $\theyt{\thefaca{},\thefaca{}}{}=0$ and
\begin{align}
\frac{\partial \exp\left({\mathbf y}_{\theconstga{},\thefaca{}}\right)}{\partial a} = 
\Delta t \, \thewt{\thefaca{}}
\exp\left(\theyt{\theconstga{},\thefaca{}}{}\right) ,    
\label{exp_001b}
\end{align}
it is straightforward to verify that eq.\,\eqref{exp_002} solves eq.~\eqref{std_002a} and is thus an alternative to eq.~\eqref{std_001a}. Since this holds for any $\theconstga$, we will set it to zero from now on. Truncating the Magnus series after the first term and applying the mean-value approximation [eq.~\eqref{mvt}], the matrix exponential becomes
\begin{align}
\exp\left(\theyt{\theconstal{},\thefaca{}}{}\right) 
&\approx 
\exp \left(
\left[\thefaca{} - \theconstal{}\right]\thedt \thewt{\thefacb{}} 
\right)
=
\theut{\thefacb{}}{\thefaca{}-\theconstal{}} ,
\label{exp_000a}
\end{align}
where we have introduced the matrix exponential
\begin{equation}
\theut{\thefacb{}}{}
= \exp(\thedt{} \thewt{\thefacb{}})
\label{exp_000b}
\end{equation}
Equation~\eqref{exp_002} now becomes:
\begin{align}
\thept{\thefaca{}} &\approx \thedt{} \int \limits_{0}^{\thefaca{}} \theut{\thefacb{}}{\thefaca{}-\theconstal{}}\mathrm{d} \theconstal \,  \theft{\thefacb{}} + \theut{\thefacb{}}{\thefaca{}} \thept{0} 
\label{exp_004}
\end{align}
Approximating the remaining integral via the midpoint rule [eq.~\eqref{mpr}], we obtain
\begin{align}
\thept{\thefaca{}} &\approx
\thefaca{} \thedt{} \theut{\thefacb{}}{\thefaca{}/2} \theft{\thefacb{}}  + \theut{\thefacb{}}{\thefaca{}} \thept{0} \nonumber \\
&= 
\theut{\thefacb{}}{\thefaca{}/2} \left[
\thefaca{} \thedt{}  \theft{\thefacb{}}  + \theut{\thefacb{}}{\thefaca{}/2} \thept{0} \right]
\label{exp_005}
\end{align}
as the exponential propagator (EXP) for the nuclear momenta in magnetic fields; see Algorithm~\ref{alg_exp}.  The matrix exponential can be expanded as
\begin{align}
\theut{\thefacb{}}{\thefacc{}} &= 
\sum \limits_{n=0}^{\infty} \dfrac{1}{n!} \left [\thefacc{} \thedt{} \thewt{\thefacb{}}\right]^{n}
\nonumber \\
&=
\sum \limits_{n=0}^{\then{}-1} \dfrac{1}{n!} \left [\thefacc{} \thedt{} \thewt{\thefacb{}}\right]^{n} + \mathcal{O} \left( 
\left [\thedt{} \thecyclofreq{} \right]^{\then{}} / \then{}!
\right) .
\label{exp_008}
\end{align}
In contrast to the Neumann series, this series converges unconditionally, for all step sizes.

\begin{algorithm}
\caption{General algorithm for the Exponential (EXP) method of order $K$. See eqs.~\eqref{eom_006a}-\eqref{eom_006d} for definitions of the variables. $\mathbf{a}$ and $\mathbf{b}$ are predefined vectors of length $K+1$ and $K$, respectively. In the velocity Verlet variant, $\mathbf{a} = [0.5,0.5]$ and $\mathbf{b} = [1.0]$.}
\begin{algorithmic}
\State $k = 0$; $a' = 0$; $b' = 0$
\While{$k \leq K$}
  \State $\thefaca{} = \mathbf{a}[k]$
  \State $\thept{a' + \thefaca{}} = \thefaca{}\thedt{} \theut{b'}{\thefaca{}/2} \theft{b'} + \theut{b'}{\thefaca{}} \thept{a'}$
  \State $a' = a' + a$
  \If{$k < K$}
    \State $\thefacb{} = \mathbf{b}[k]$
    \State $\thext{b' + \thefacb{}} = \thefacb{}\thedt{} \left[ \them{}^{-1} \thept{a'} \right] + \thext{b'}$
    \State $b' = b' + b$
    \State \textbf{Calculate} $\theft{b'}$ and $\thewt{b'}$
  \EndIf
  \State $k = k+1$
\EndWhile
\end{algorithmic}
\label{alg_exp}
\par\noindent\rule{0.48\textwidth}{0.1pt}
\end{algorithm}

\subsection{Comparison of the propagators}

We close this section by discussing a few properties of the introduced propagators. The results are summarized in Table\,\ref{tab_comp}. From Algorithms~\ref{alg_aux}--\ref{alg_exp}, we see that the ACM propagator is about three times more expensive than the other schemes, requiring $3K$ instead of $K$ forces and Berry curvature calculations per step. In addition, it depends on the definition of a parameter ($\thecoupl{}$), which has an influence on the stability of the dynamics. Moreover, unlike the ACM propagator, TAJ and EXP propagators converge to the correct zero-field solution when setting $\theb{}$ and therefore $\thewt{b}$ to zero.

As the correct zero-shielding limit, we define the propagators derived by Spreiter and Walter, which were constructed for the special case where the Berry curvature is zero. While their working equations clearly differ from the ACM propagator, we can compare them to the TAJ and EXP propagators by assuming a single particle with charge $\thez{1}$ and mass $\them{1}$, experiencing a time-dependent external force $\theft{}$ and a magnetic field of strength $\theb{z}$ in the z-direction. In this particular case, $\thewt{}$ and thus $\theut{}{a}$ and $\thevt{}{a}$ are constants
\begin{align}
\thewt{} =
\begin{pmatrix} 
0 & \thecyclofreq{}  & 0  \\
-\thecyclofreq{}  & 0 & 0 \\
0 & 0 & 0
\end{pmatrix} \qquad
\thecyclofreq{} = \dfrac{\theb{z} \thez{1}}{\them{1}} ,
\label{com_000}
\end{align}
and the velocity Verlet variants of the EXP and TAJ propagator reduce to
\begin{align}
\thept{1}^\mathrm{EXP} &=
\dfrac{\thedt{}}{2} \left(\theut{}{0.25} \theft{1}  + 
 \theut{}{0.75} \theft{0} \right) + 
\theut{}{1} \thept{0} ,
\label{com_001a} \\
\thept{1}^\mathrm{TAJ} &=
\dfrac{\thedt{}}{2} \left(\thevt{}{0.25} \theft{1}  + 
\thevt{}{0.25} 
\left[\theone + \dfrac{\thedt{}}{4} \thewt{}\right]
\thevt{}{0.25} \theft{0} \right) \nonumber \\
&\quad+
\thevt{}{0.25} 
\left[\theone + \dfrac{\thedt{}}{4} \thewt{}\right]
\thevt{}{0.25}
\left[\theone + \dfrac{\thedt{}}{4} \thewt{}\right]
\thept{0} ,
\label{com_001b}
\end{align}
respectively. Expanding $\theut{}{a}$ and $\thevt{}{a}$ up to second order
\begin{align}
\theut{}{\thefaca{}} &= \theone + \thefaca{} \thedt{} \thewt{} + 
\dfrac{1}{2} \thefaca{}^2 (\thedt{})^2 \thewt{}^2 + \mathcal{O} \left([\thedt{}]^3\right) 
\\
\thevt{}{\thefaca{}} &= \theone + \thefaca{} \thedt{} \thewt{} + 
\thefaca{}^2 (\thedt{})^2 \thewt{}^2 + \mathcal{O} \left([\thedt{}]^3\right) 
\label{com_001c}
\end{align}
and using the Taylor expansion of the forces in y-direction
\begin{align}
\theftel{1}{y} = \theftel{0}{y} + \thedt{} \dfrac{d \theftel{0}{y}}{dt} + \mathcal{O} \left([\thedt{}]^2\right) ,
\label{com_002}
\end{align}
we obtain, for both EXP and TAJ, an expression that is identical to those obtained by Spreiter and Walter via inversion [see eq.~(16) in ref.~\onlinecite{Spreiter1999a}] and via Taylor expansion [see eq.~(39) in ref.~\onlinecite{Spreiter1999a}]:
\begin{align}
\theptel{1}{x} &=
\theptel{0}{x} + 
\dfrac{\thedt{}}{2} \left[
\theftel{0}{x} + \theftel{1}{x} + 2 \thecyclofreq{} \theptel{0}{y}
\right] \nonumber \\&\quad+ 
\dfrac{1}{4}
\left(\thedt\right)^2
\left[
2 \thecyclofreq{} \theftel{0}{y}
- 2 \thecyclofreq{}^2 \theptel{0}{x} 
\right]
+ \mathcal{O} \left([\thedt{}]^3\right) 
\label{com_003}
\end{align}

Their difference, however, becomes obvious when additionally setting the forces in eqs.~\eqref{com_001a} and \eqref{com_001b} to zero:
\begin{align}
\thept{1}^\mathrm{EXP} &=
\theut{}{1} \thept{0} 
\label{exp_007} \\
\thept{1}^\mathrm{TAJ} &=
\thevt{}{0.25} 
\left[\theone + \dfrac{\thedt{}}{4} \thewt{}\right]
\thevt{}{0.25}
\left[\theone + \dfrac{\thedt{}}{4} \thewt{}\right]
\thept{0} 
\label{exp_007b}
\end{align}
In this \emph{cyclotron} limit, the EXP propagator collapses to the exact result for the cyclotron motion of a single, charged particle
\begin{align}
\thept{1}^\mathrm{EXP} &=
\begin{pmatrix} 
\cos \left(\thedt{} \thecyclofreq{} \right) & \sin \left(\thedt{} \thecyclofreq{} \right)  & 0  \\
-\sin \left(\thedt{} \thecyclofreq{} \right) & \cos \left(\thedt{} \thecyclofreq{} \right) & 0 \\
0 & 0 & 0
\end{pmatrix}
\thept{0} ,
\label{exp_007c} 
\end{align}
while TAJ introduces an error at the order of $\mathcal{O} ( \left [\thedt{} \thecyclofreq{} \right]^{3} )$. For the previously mentioned estimate [see eq.~\eqref{dt_003}] with $B = 0.1$ and $\thedt{} = 10$, this results in an error at the order of $10^{-12}$ (all given in atomic units). Thus, in practice EXP and TAJ trajectories and their computational cost will be very similar.

\begin{table}
\centering
\caption{Comparison of the auxiliary coordinates and momenta (ACM), Tajima (TAJ), and exponential (EXP) propagator.}
\begin{tabular}{c|ccc}
Criterion & ACM & TAJ & EXP \\ \hline
Forces calculations per step & $3K$ & $K$ & $K$ \\
Parameter-free? & No & Yes & Yes \\
Correct zero-field limit? & No & Yes & Yes \\
Correct zero-shielding, & \multirow{2}{*}{No} & \multirow{2}{*}{Yes} & \multirow{2}{*}{Yes} \\
velocity Verlet limit? & \\
Exact cyclotron limit? & No & No & Yes
\end{tabular}
\label{tab_comp}
\par\noindent\rule{0.48\textwidth}{0.1pt}
\end{table}

\section{Computational Details}

\subsection{Diatomic model system}

\begin{figure*}
\centering
\begin{tabular}{ll}
(a) & (b) \\
\includegraphics[width=0.48\textwidth]{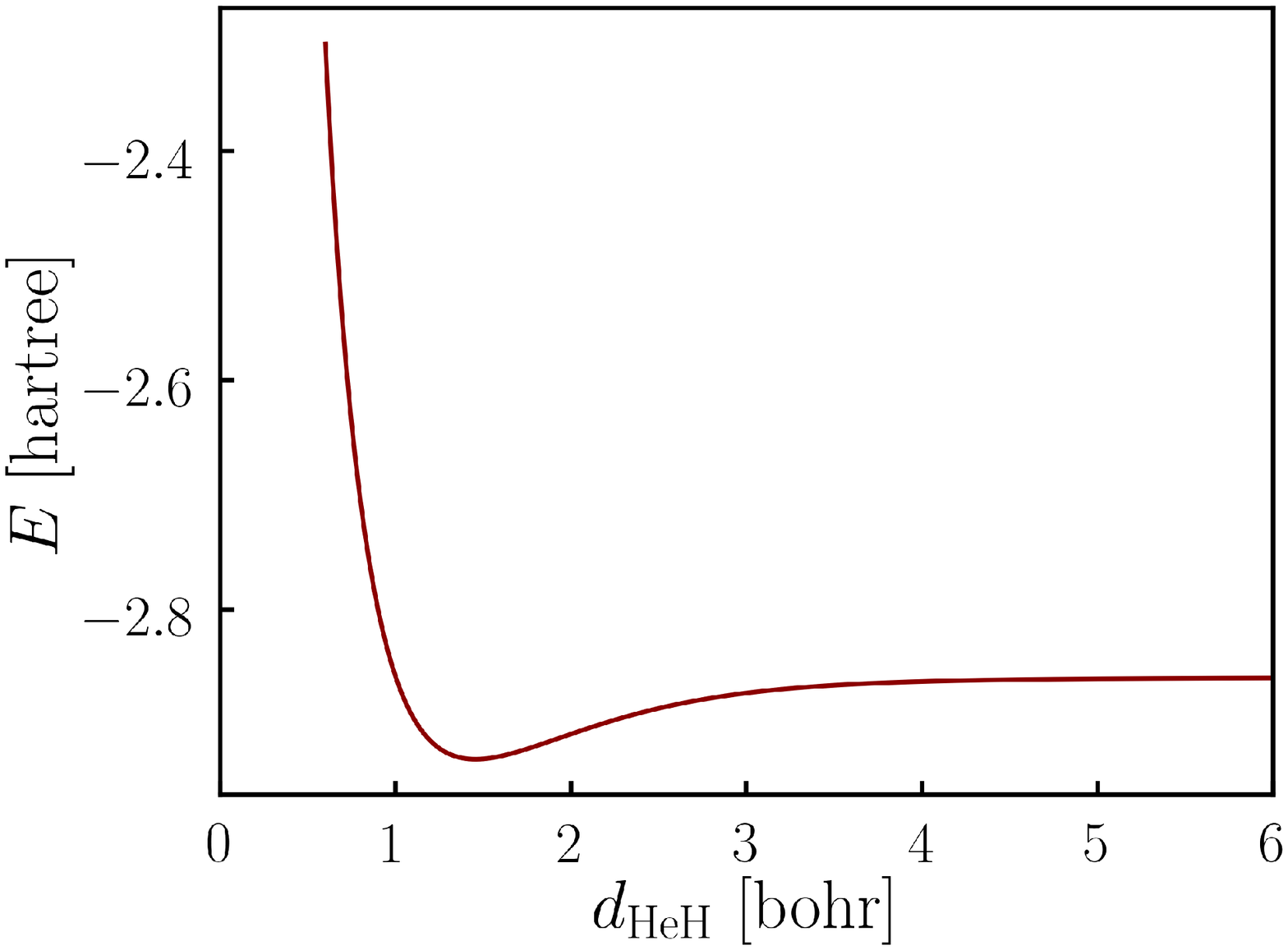}  &  
\includegraphics[width=0.48\textwidth]{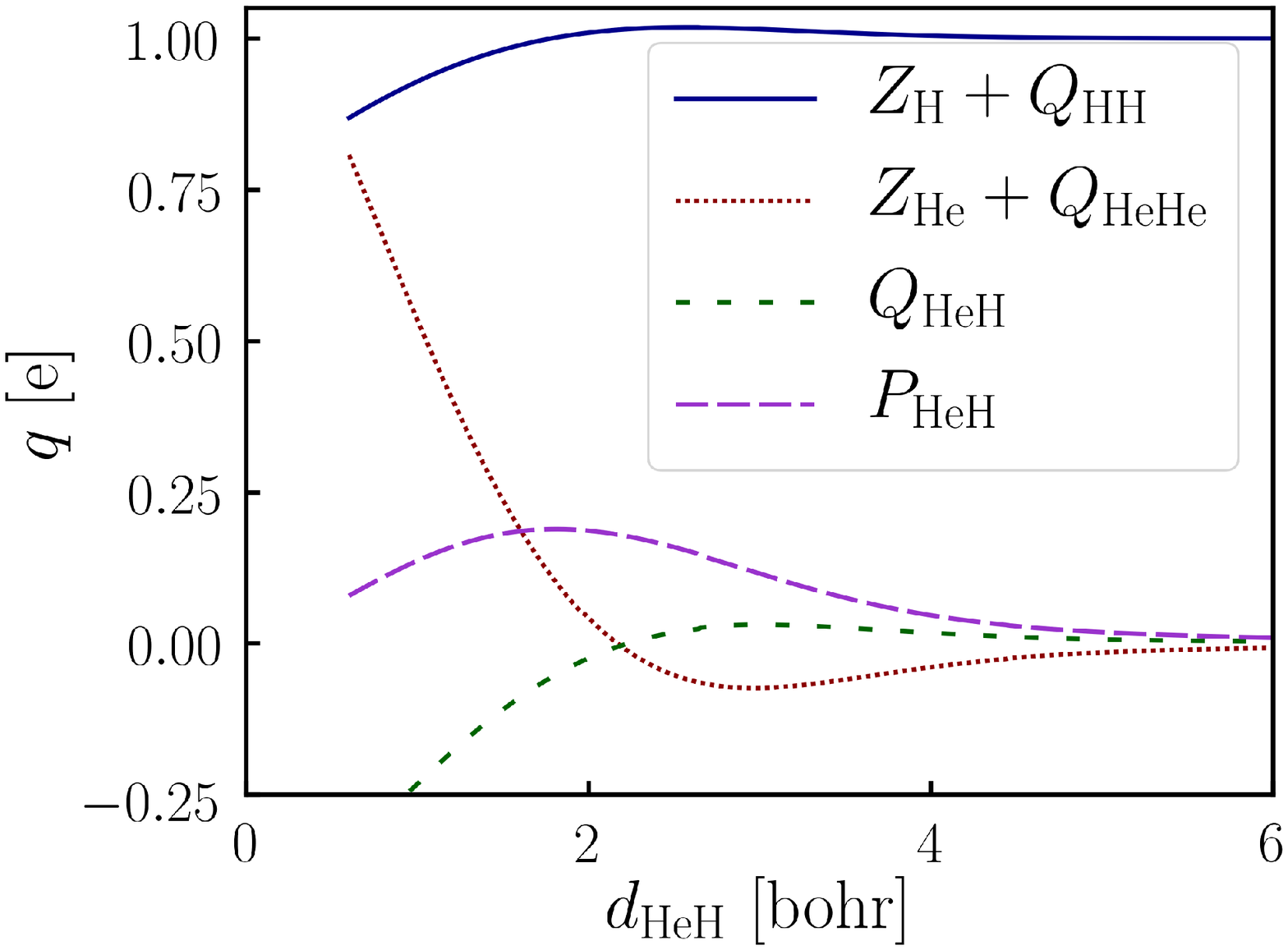}
\end{tabular}
\caption{Bond-length ($d_\mathrm{HeH}$) dependent energies (a) and charges (b) derived from the Berry curvature of HeH$^+$ at the HF/lu-aug-cc-pVTZ level of theory perpendicular to a field of $B=0.1\,$B$_0$. We consistently use atomic units.}
\label{fig_q}
\par\noindent\rule{0.98\textwidth}{0.1pt}
\end{figure*}

To conduct a study of the performance of the previously discussed propagators, we carry out simulations of a \emph{diatomic model system} perpendicular to the magnetic field. It has been shown that, in this particular case, the Berry curvature depends solely on the Berry charges $\theq{\indnuc}{\indnuctwo}{\ther{},\theb{}}$ and Berry charge fluctuations $\thep{\indnuc}{\indnuctwo}{\ther{},\theb{}}$ as well as the orientation of the molecule relative to the field vector:\cite{Culpitt2021b,Peters2023a}
\begin{align}
\theom{\indnuc}{\indnuctwo}{\ther{}, \theb{}} &= - \theq{\indnuc}{\indnuctwo}{\ther{},\theb{}} \thebt{} \nonumber \\
&\quad - \thep{\indnuc}{\indnuctwo}{\ther{},\theb{}} 
\left[
\thebt{} \bar{\mathbf{R}}_{\indnuc\indnuctwo} \bar{\mathbf{R}}_{\indnuc\indnuctwo}^\mathrm{T} - \bar{\mathbf{R}}_{\indnuc\indnuctwo}\bar{\mathbf{R}}_{\indnuc\indnuctwo}^\mathrm{T} \thebt{}
\right] ,
\label{comp_000}
\end{align}
where $\bar{\mathbf{R}}_{\indnuc\indnuctwo}$ is the normalized interatomic distance vector:
\begin{align}
\bar{\mathbf{R}}_{\indnuc\indnuctwo} = d_{\indnuc\indnuctwo}^{-1} \left[\ther{\indnuctwo} - \ther{\indnuc}\right]
\qquad
d_{\indnuc\indnuctwo} = |\ther{\indnuctwo} - \ther{\indnuc}|
\label{comp_001}
\end{align}
Our model system has been designed to reproduce the properties of HeH$^+$ at field strength $0.1\,\mathrm{B}_0$, with energies $E (d_\mathrm{HeH})$, Berry charges $\theq{\indnuc}{\indnuctwo}{d_\mathrm{HeH}}$, and Berry charge fluctuations $\thep{\indnuc}{\indnuctwo}{d_\mathrm{HeH}}$ obtained by spline fitting of Hartree--Fock (HF)/lu-aug-cc-pVTZ\cite{Dunning1989b,Kendall1992,Woon1994} results for different bond lengths ($d_\mathrm{HeH}$) at this field strength. The prefix \enquote{lu-} indicates the use of an uncontracted London orbital basis set, as suggested in ref.~\onlinecite{Tellgren2008b}. All calculations were performed with the \textsc{London}\cite{London} program package. The resulting bond-length dependence of the energy and charges is shown in Fig.\,\ref{fig_q}. We note that HeH$^+$ dissociates into He with effective charge $Z_\text{He}+ Q_\text{HeHe} = 0$ and a proton H$^+$ with effective charge $Z_\text{H}+ Q_\text{HH} = 1$. In the bonding region, both nuclear charges are partially screened by the electrons. 

Note that, in our calculations, we neglect the magnetic-field dependence of the electronic energy itself $E (d_\mathrm{HeH})$, of the Berry charges, $\theq{\indnuc}{\indnuctwo}{d_\mathrm{HeH}}$, and
of the Berry fluctuations $\thep{\indnuc}{\indnuctwo}{d_\mathrm{HeH}}$. The magnetic field therefore enters the calculations only through
the explicit field-dependence of the model Berry curvature---that is,
by the dependence on $\tilde {\mathbf B}$ in eq.\,\eqref{comp_000}. This approach allows for a more systematic study of the propagators, since only the velocity-dependent forces increase linearly with $B$, while the potential-energy surface and the electronic structure remain unaffected. Because of its overall charge, small mass, and significant geometry dependence of the screening [see Fig.~\ref{fig_q}(b)], the HeH$^+$ model system can be regarded as an extreme case, featuring (comparably) large cyclotron frequencies ($\thecyclofreq{}$). 

\subsection{Molecular simulations}

All NVE simulations (constant number of particles, volume, and energy) were conducted in the $xy$-plane with the magnetic field aligned along the $z$-axis. Each simulation began at the equilibrium geometry, with random initial momenta. We investigated the ACM propagator (Algorithm~\ref{alg_aux}), the TAJ propagator (Algorithm~\ref{alg_taj}), and the exponential operator (Algorithm~\ref{alg_exp}) at three different orders $K$: the velocity-Verlet (VV) scheme\cite{Verlet1967a,Swope1982a} with $K=1$, the OM scheme of Omelyan \emph{et al.}\cite{Omelyan2003a} with $K=4$, and the RK4 scheme (S$_6$/O4 in ref.\,\onlinecite{Blanes2002a}) with $K=6$. The corresponding factors $\mathbf{a}$ and $\mathbf{b}$ are given in Tab.~\ref{tab_ab}.

Each trajectory was simulated for 24$\,$ps using an effective time step ($\thedt{}/K$) of 0.1$\,$fs, storing energies, geometries, and momenta every 2.4$\,$fs. For the ACM propagator, we used an optimized coupling constant $\thecoupl{}=0.013$. Where $\then{}$ is not given, we used the standard matrix-inversion and exponential algorithms of \texttt{python3.6} to calculate $\thevt{}{}$ and $\theut{}{}$ in the TAJ and EXP propagators, respectively, and the truncated series of eqs.~\eqref{taj_004} and \eqref{exp_008} otherwise. Estimating that $\thedt\,\thecyclofreq{}=4.5\times10^{-4}B$, we apply a magnetic-field range of $10^{-2}$--$10^{3}$ in units of $B_0$ to ensure that the weak-field condition in eq.~\eqref{dt_002} holds for all simulations.

We use the standard deviation of the total energy ($\varepsilon_\mathrm{tot}$) as a criterion for the stability of the dynamics. In addition, vibrational spectra\cite{Thomas2013,Peters2021b} and the center-of-mass motion are used to compare trajectories of the different propagators. 

\begin{table}[ht]
\caption{Coefficients $\mathbf{a}$ and $\mathbf{b}$ for the different integrators of order $K$ used in this work.}
\label{the_tab}
\begin{tabular}{c|c}\hline\hline
\multicolumn{2}{c}{VV ($K = 1$)} \\
$a_{0}$ = $\phantom{-}$0.5000000000000000 & $b_{0}$ = $\phantom{-}$1.0000000000000000 \\
$a_{1}$ = $\phantom{-}$0.5000000000000000 & \\ \hline
\multicolumn{2}{c}{OM ($K = 4$)} \\
$a_{0}$ = $\phantom{-}$0.1786178958448091 & $b_{0}$ = $\phantom{-}$0.7123418310626054 \\
$a_{1}$ = $-$0.0662645826698185 & $b_{1}$ = $-$0.2123418310626054 \\
$a_{2}$ = $\phantom{-}$0.7752933736500187 & $b_{2}$ = $-$0.2123418310626054 \\
$a_{3}$ = $-$0.0662645826698185 & $b_{3}$ = $\phantom{-}$0.7123418310626054 \\
$a_{4}$ = $\phantom{-}$0.1786178958448091 & \\ \hline
\multicolumn{2}{c}{RK4 ($K = 6$)} \\
$a_{0}$ = $\phantom{-}$0.0792036964311957 & $b_{0}$ = $\phantom{-}$0.2095151066133620 \\
$a_{1}$ = $\phantom{-}$0.3531729060497740 & $b_{1}$ = $-$0.1438517731798180 \\
$a_{2}$ = $-$0.0420650803577195 & $b_{2}$ = $\phantom{-}$0.4343366665664560 \\
$a_{3}$ = $\phantom{-}$0.2193769557534996 & $b_{3}$ = $\phantom{-}$0.4343366665664560 \\
$a_{4}$ = $-$0.0420650803577195 & $b_{4}$ = $-$0.1438517731798180 \\
$a_{5}$ = $\phantom{-}$0.3531729060497740 & $b_{5}$ = $\phantom{-}$0.2095151066133620 \\
$a_{6}$ = $\phantom{-}$0.0792036964311957 & \\ 
 \end{tabular}
 \label{tab_ab}
 \par\noindent\rule{0.48\textwidth}{0.1pt}
\end{table}

\section{Results and Discussion}

\begin{figure}
\centering
\includegraphics[width=0.48\textwidth]{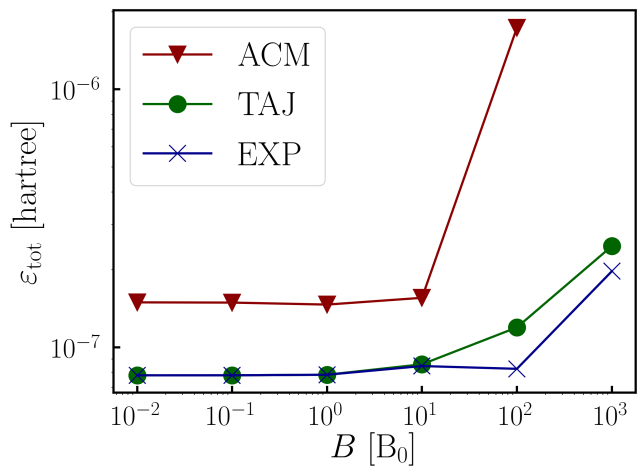} 
\caption{Standard deviation of the total energy ($\varepsilon_\mathrm{tot}$) during simulations of the HeH$^+$ model system using different magnetic field strengths and propagators with $K = 1$ (velocity Verlet): Auxiliary coordinates and momenta (ACM), Tajima (TAJ), and exponential (EXP) propagator.}
\label{fig_comp}
\par\noindent\rule{0.48\textwidth}{0.1pt}
\end{figure}

In Fig.~\ref{fig_comp}, we have plotted $\varepsilon_\mathrm{tot}$ for HeH$^+$ as a function of field strength $B$ for the ACM, TAJ, and EXP propagators with $K=1$. The results for $K=4$ and $K=6$ are very similar and therefore not included in the figure.

In general, all three propagators become less stable with increasing field strength---in particular the ACM propagator, for which the simulation at $10^{3}\,\mathrm{B}_0$ becomes unstable. At this high field strength (characteristic of neutron stars rather than white dwarfs), a reoptimization of the coupling-strength parameter $s$ may be required. In agreement with a previous comparison of the ACM and TAJ propagators,\cite{Monzel2022b} we note that the ACM propagator performs less well than the TAJ and EXP propagators. Since, in addition, the TAJ and EXP propagators are parameter-free and since they require only one rather than three force and Berry-curvature calculations per step, we conclude that the the TAJ and EXP propagators are to be preferred over the ACM propagator.

\begin{figure*}
\centering
\begin{tabular}{ll}
(a) & (b) \\
\includegraphics[width=0.48\textwidth]{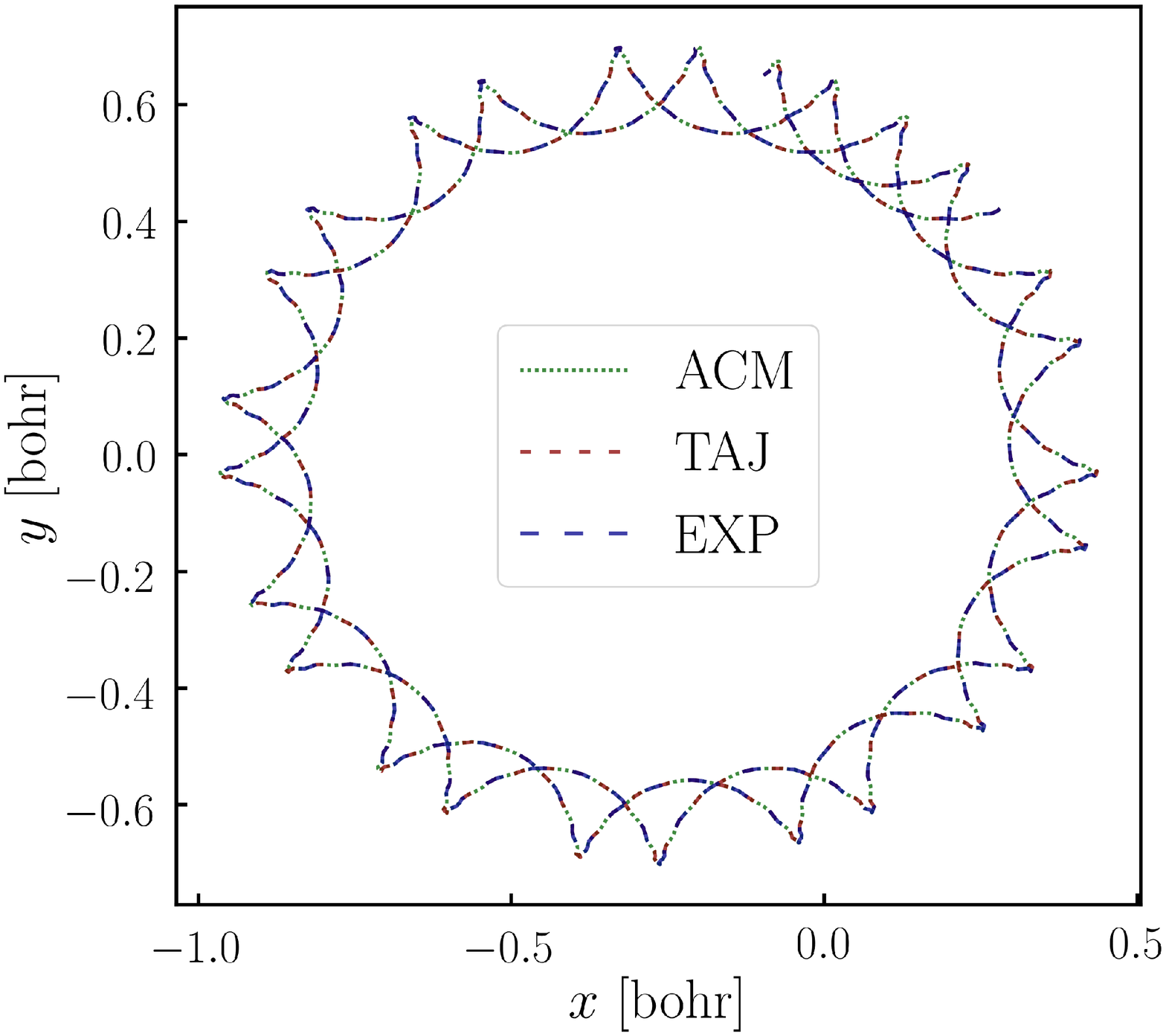} &
\includegraphics[width=0.48\textwidth]{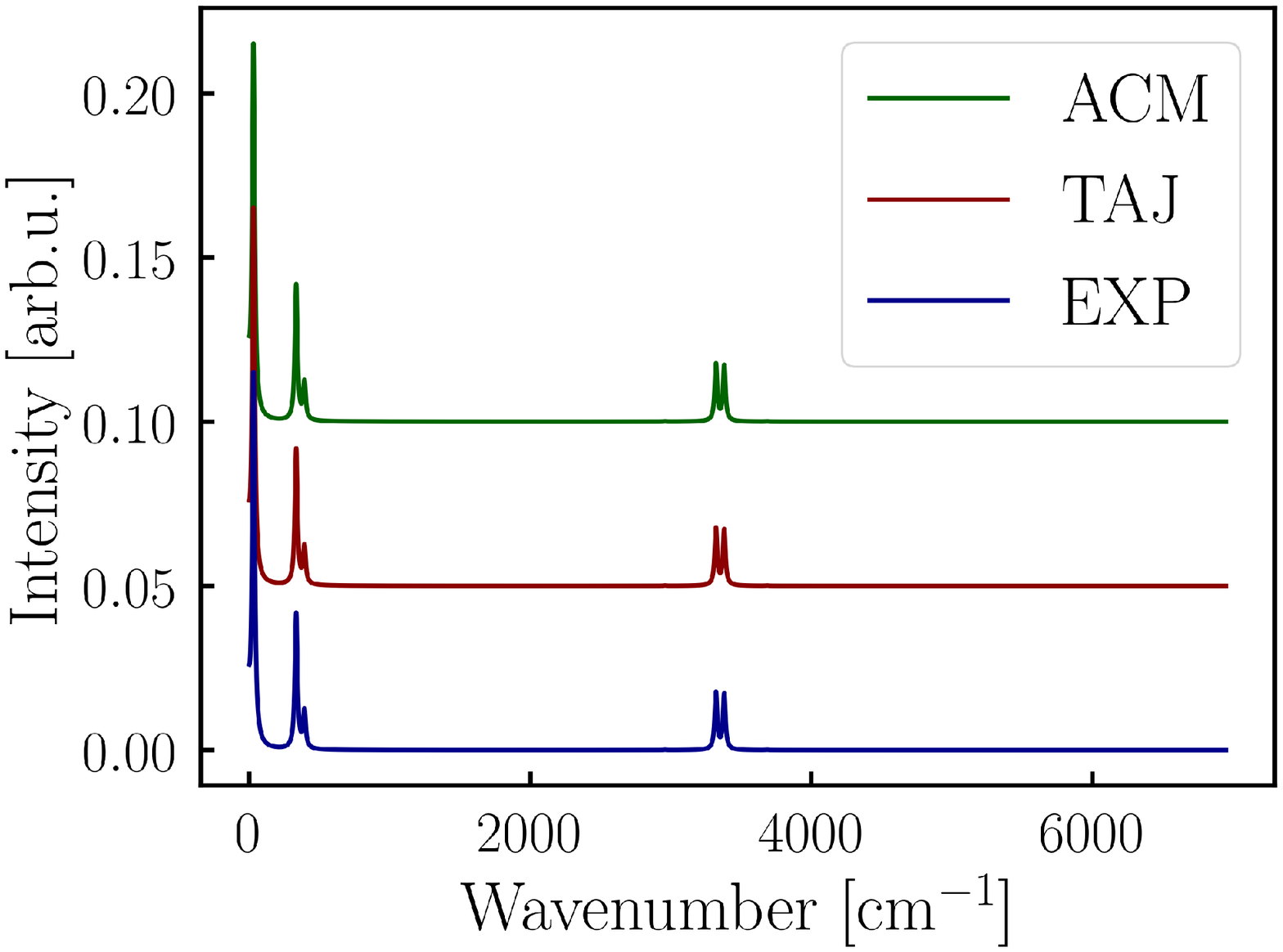} 
\end{tabular}
\caption{Center-of-mass motion (a) and vibrational spectra of He (b) obtained from the simulations of the HeH$^+$ model system at $B = 10\,$B$_0$ using different propagators with $K = 1$ (Velocity Verlet): Auxiliary coordinates and momenta (ACM), Tajima (TAJ), and exponential (EXP) propagator.}
\label{fig_traj}
\par\noindent\rule{0.98\textwidth}{0.1pt}
\end{figure*}

While the EXP and TAJ propagators are indistinguishable at low field strengths, the EXP propagator becomes marginally more stable for $B > 10\,\mathrm{B}_0$, as the cyclotronic center-of-mass motion of the charged HeH$^+$ system begins to dominate the trajectories. However, even at $10\,\mathrm{B}_0$, the center-of-mass motions and the vibrational spectra of the two propagators are still indistinguishable and in fact identical to those obtained with the ACM propagator; see Fig.~\ref{fig_traj}. In line with previous studies on diatomics,\cite{Peters2021b,Monzel2022b} the peaks in the vibrational spectrum of HeH$^+$ (Fig.~\ref{fig_traj}b) correspond to the cyclotronic center-of-mass motion ($\sim30\,$cm$^{-1}$, also visible in Fig.~\ref{fig_traj}a), rotation ($\sim370\,$cm$^{-1}$), and vibration ($\sim3350\,$cm$^{-1}$), and feature splitting patterns that arise from the couplings between the different modes.

\begin{figure*}
\centering
\begin{tabular}{ll}
(a) & (b) \\
\includegraphics[width=0.48\textwidth]{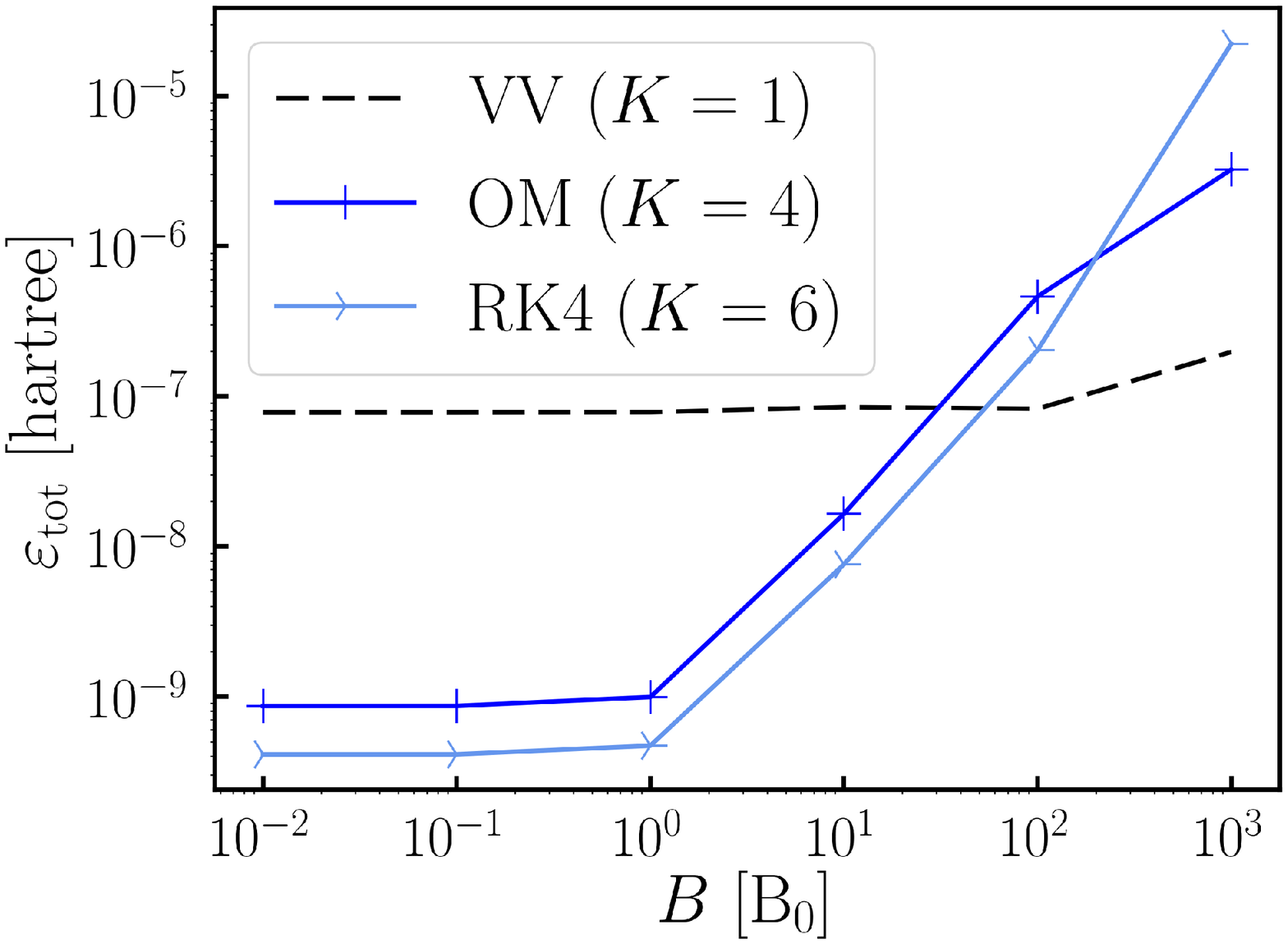} &
\includegraphics[width=0.48\textwidth]{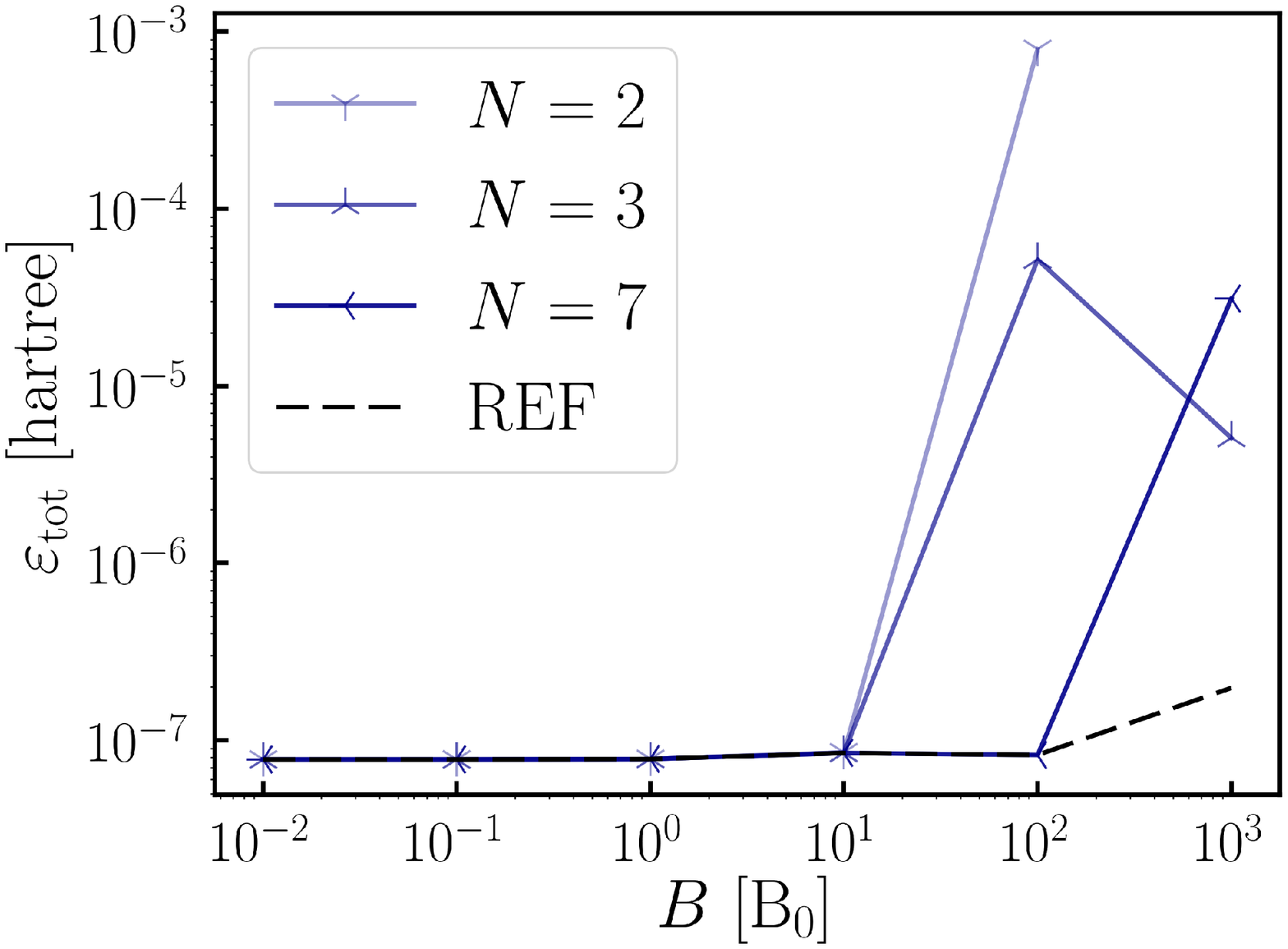} 
\end{tabular}
\caption{Influence of the order of the EXP propagator ($K$, a) and the truncation of the exponential series ($N$, b) on the standard deviation of the total energy ($\varepsilon_\mathrm{tot}$) during simulations of the HeH$^+$ model system using different magnetic field strengths. The reference (exact exponential with $K = 1$) is shown as a dashed black line in both plots.}
\label{fig_mn}
\par\noindent\rule{0.98\textwidth}{0.1pt}
\end{figure*}

Since the EXP propagator is the most stable propagator, we restrict our attention to this propagator when investigating dependence on the propagator order $K$ and on the truncation level $N$ in the exponential series; see Fig.\,\ref{fig_mn}(a) and (b), respectively. We expect the TAJ propagator to show a similar behaviour, while an  investigation of the influence of $K$ on the ACM propagator can be found in ref.\,\onlinecite{Peters2021b}.

At low field strengths, the higher-order schemes (i.e., OM with $K=4$ and RK4 with $K=6$) improve the stability of the EXP propagator by up to two orders of magnitude; see Fig.~\ref{fig_mn}. These higher-order schemes therefore allow for a significantly larger effective time step $\thedt{}/K$ (constant in Fig.~\ref{fig_mn}), reducing the computational cost of the dynamics. Perhaps surprisingly, the OM and RK4 schemes become less stable than the lower-order VV scheme as the velocity-dependent forces become dominant in the Landau regime (beyond $10\,\mathrm{B}_0$). This might be due to the fact that the coefficients $\mathbf{a}$ and $\mathbf{b}$ (see Algorithm~\ref{alg_exp}) were not optimized for such propagations.

Simulations with a truncated exponential series at $N=2$ in eq.~\eqref{exp_008} reproduce the results obtained with the exact matrix exponential for field strengths up $10\,\mathrm{B}_0$; see Fig.~\ref{fig_mn}. Since calculating a matrix exponential is (comparably) expensive, replacing it with a series will reduce the computation time. For $N=2$ at higher field strengths, the truncation error exceeds $\mathcal{O} \left([\thecyclofreq{}\thedt{}]^2\right) = 10^{-5}$, giving unstable dynamics. This indicates that $N$ must be chosen with care.

\section{Conclusion and Outlook}

In this work, we have studied three propagators for molecular dynamics in a magnetic field---namely, the auxiliary-coordinates-and-momenta (ACM) propagator, the Tajima (TAJ) propagator, and the new exponential (EXP) propagator, testing their performance using simulations of a HeH$^+$ model system at different field strengths ($10^{-2}$-- $10^{3}\,\mathrm{B}_0$). 

The EXP propagator, which was derived from a truncated Magnus expansion, correctly reduces to the standard velocity Verlet propagator in the zero-field limit, the propagators of Spreiter and Walter\cite{Spreiter1999a}, neglecting electron shielding, and the cyclotronic motion of a charged particle. Since it also performed best in our model simulations and showed the best stability, especially at higher field strengths, we recommend the EXP propagator for simulations of molecules in a magnetic field. However, the TAJ propagator delivers identical results at field strengths $\leq 1\,\mathrm{B}_0$ and the more expensive ACM propagator might be useful for cases where the energy and/or the electron shielding depend on the nuclear momenta. 

In the study of molecules in strong magnetic fields, we are usually interested in field strengths below $1\,\mathrm{B}_0$ and apply time steps of $\thedt=0.1\,$fs to resolve the molecular vibrations. In practice, these time steps are small enough to resolve the cyclotronic motion (\emph{weak field limit}), while, additionally, allowing for the use of higher-order schemes and truncation of the exponential series to reduce the computational cost of the EXP propagator. 

\clearpage
\section*{Acknowledgements}

We thank Tanner Culpitt for helpful discussions.

\section*{Disclosure statement}

No potential conflict of interest was reported by the author(s).

\section*{Funding}

This work was supported by the Research Council of Norway through ‘‘Magnetic Chemistry’’ Grant No.\,287950 and CoE Hylleraas Centre for Quantum Molecular Sciences Grant No.\,262695. The work also received support from the UNINETT Sigma2, the National Infrastructure for High Performance Computing and Data Storage, through a grant of computer time (Grant No.\,NN4654K).

\section*{References}

%\bibliography{prop_in_b_new,lit}

\end{document}